\documentclass[useAMS,twocolumn,usenatbib]{mn2e}
\usepackage{amssymb}
\usepackage{amsmath}
\usepackage{graphicx}
\DeclareGraphicsExtensions{.eps, .ps, .eps.gz, ps.gz}
\usepackage[british]{babel} 
\usepackage{placeins}
\usepackage{subfigure}
\usepackage{hyperref}
\usepackage{color}

\newcommand{\rev}[1]{{#1}}

\newcommand{\paperII}{paper I}
\newcommand{\paperIII}{paper II}

\newcommand{\Msun}{{$h^{-1}M_\odot$}}
\newcommand{\Mpc}{{$h^{-1}\text{Mpc}$}}

\newcommand{\dif}{{\mathrm d}}
\newcommand{\bxicross}{$b_{\xi}^{\times}$ }
\newcommand{\bdeltaq}{$b_{\Delta Q}$ }
\newcommand{\btaucross}{$b_{\tau}^{\times}$ }

\begin{document}


\title[bias predictions vs. measurements]{
Linear and non-linear bias: predictions vs. measurements
}
\author[K. Hoffmann, J. Bel, E. Gazta\~naga] 
{K. Hoffmann$^{1,2}$, J. Bel$^{3}$, E. Gazta\~naga$^{1}$ \\
$^{1}$Institut de Ci\`{e}ncies de l'Espai (ICE, IEEC/CSIC), E-08193 Bellaterra (Barcelona), Spain \\
$^{2}$Center for Astrophysics, Department of Physics, Tsinghua University, Beijing, 100084, China \\
$^{3}$Aix Marseille Univ, Univ Toulon, CNRS, CPT, Marseille, France\\
} 

\date{Received date / Accepted date}

\maketitle

\begin{abstract}
We study the linear and non-linear bias parameters which determine the mapping between the distributions of
galaxies and the full matter density fields, comparing different measurements and predictions. Associating galaxies with dark matter haloes
in the MICE Grand Challenge N-body simulation we directly measure the bias
parameters by comparing the smoothed density fluctuations of haloes and matter in the same region at
different positions as a function of smoothing scale. Alternatively we measure the bias parameters 
by matching the probability distributions of halo and matter density fluctuations, which can be applied to observations. These direct
bias measurements are compared to corresponding measurements from two-point and different third-order
correlations, as well as predictions from the peak-background model, which we presented in previous
articles using the same data. We find an overall variation of the linear bias measurements and predictions of $\sim 5 \%$ with
respect to results from two-point correlations for different halo samples with masses between
$\sim 10^{12} - 10^{15}$ $h^{-1}M_\odot$ at the redshifts $z=0.0$ and $0.5$. Variations between
the second- and third-order bias parameters from the different methods show larger variations, but
with consistent trends in mass and redshift.
The various bias measurements reveal a tight relation between the linear and the quadratic bias 
parameters, which is consistent with results from the literature based on simulations with  different cosmologies.
Such a universal relation might improve constraints on cosmological models, derived from second-order 
clustering statistics at small scales or higher-order clustering statistics.
\end{abstract}


\section{Introduction}\label{sec:introduction}

The increase of data from upcoming and next generation of galaxy surveys
is pulling down errors on the observed statistics of the large-scale galaxy distribution.
Thus, the inferences on cosmological models from these statistics requires a
modelling of cosmological fluids and their statistical properties with an accuracy of at least the
same order of magnitude as the observational errors.

One of the largest uncertainties comes from the modelling of the mapping between the observed fluctuations 
of galaxies to the fluctuation of the underlying matter distribution (hereafter referred to as $\delta_g$ and
$\delta_m$ respectively). These fluctuations are defined as normalised
deviations of the density $\rho$, smoothed typically with a top-hat window of characteristic scale 
$R$,  from the mean density of the universe $\bar \rho$  at the position $\bf 
r$,

\begin{equation}
  \delta(\bf r) \equiv \frac{\rho(\bf r) - \bar \rho}{\bar \rho}.
  \label{eq:delta}
\end{equation}
The mapping from $\delta_g$ to $\delta_m$ is described by the
so-called bias function, $\delta_g=F[\delta_{m}, \nabla_{ij}\Phi_v]$, where $\nabla_{ij}\Phi_v$ are second-order derivatives of
the velocity potential. The latter relate $\delta_g$ to the matter distribution beyond the smoothing scale 
$R$ and are therefore referred as {\it non-local} contributions to the bias model \citep{chan12, baldauf12}.
For sufficiently large smoothing scales such non-local contributions
may be negligible, which allows for a local description of biasing. Thus,
the bias function can be modelled as a Taylor expansion in terms of the matter fluctuations
\citep{FG}
%
\begin{equation} 
\delta_h=F[\delta_{m}] \simeq \sum_{i=0}^{N}\frac{b_i}{i!}\delta_{m}^i,
\label{eq:biasmodel_local}
\end{equation}
where $b_i$ are the bias parameters. However, for smaller smoothing scales, which are typically used for studying
the two- and three-point statistics of the galaxy distribution, non-local contributions need to be 
considered. When fluctuations in the density and velocity potential of the matter distribution are sufficiently
small the non-local bias model may be described by its second-order expansion, which is 
hereafter referred to as non-local quadratic bias model

\begin{equation}
\delta_g = b_1  \biggl\{ \delta_m + \frac{1}{2}[ c_2(\delta_m^2 - \langle \delta_m^2  \rangle) + g_2  \mathcal{G}_2] \biggr\},
\label{eq:biasmodel_quadratic_nonlocal}
\end{equation}
where $b_1$ and $c_2 \equiv b_2/b_1$ are referred to as linear and quadratic (or second-order) bias 
parameters respectively. The non-local contribution to the bias function consists in that case of the
product of the second-order non-local bias parameter $g_2$ and the smoothed second-order Gallileon

\begin{equation}
  \mathcal{ G}_2({\bf r})= \int \beta_{12}\theta_v({\bf k}_1) \theta_v({\bf k}_2) \
  \hat W[k_{12}R]e^{i {\bf k}_{12}\cdot {\bf r} }d^3 {\bf k}_1 d^3 {\bf k}_2,
\label{eq:G2}
\end{equation}
where ${\bf k}_i$ and ${\bf k}_{12} \equiv {\bf k}_2 - {\bf k}_1$ are wave vectors of density oscillations,
$\beta_{12} \equiv 1 - ({\hat{\bf k}}_1 \cdot {\hat{\bf k}}_2)^2$ represents the mode-coupling between
density oscillations which describes tidal forces, $\theta_v \equiv \nabla^2 \Phi_v$ is the divergence of the
normalised velocity field (${\bf v}/\mathcal{H}/f$) and  $ \hat W[k_{12}R]$ is the window function in Fourier space. 
Note that in the case $g_2=0$ equation (\ref{eq:biasmodel_quadratic_nonlocal}) corresponds to the local quadratic bias model.

The bias parameters are highly relevant for constraining cosmological models via the growth of matter 
fluctuations, derived from second-order statistics of the observed galaxy distribution. In 
particular at large scales, the linear bias factor $b_1$ is completely degenerate with the 
linear growth factor. Hence, growth-independent measurements of $b_1$ can strongly tighten cosmological
constraints from galaxy surveys.

Third-order statistics probes the linear and quadratic bias parameters independently of the growth and can
 be used to  break the growth-bias degeneracy in the second-order statistics. Furthermore, the
second-order bias measurements from third-order statistics allow growth measurements from
second-order statistics at small scales, where non-linear and non-local terms contribute significantly to the signal.
Such a small scale analysis would strongly improve the cosmological constraining power of galaxy surveys.

However, the value of combining second- with third-order statistics for constraining cosmological models
with high precision arises from a detailed understanding of how exactly the bias parameters enter 
these statistics at different scales, redshifts and for different
samples of tracers.

We therefore investigated in previous works
\citep[][where the latter is hereafter referred to as \paperII]{HBG15-1, BHG15} how accurately the linear bias can be
measured from different third-order statistics. These studies were based on the large cosmological
MICE Grand Challenge (hereafter referred to as MICE-GC) N-body simulation in which haloes were detected as tracers of the cosmic web
and associated with galaxies. The fact that the dark matter distribution is accessible in simulations allows for
reliable measurements of the linear bias via second-order statistics, which can then be used as a reference
for validating linear bias measurements from third-order statistics as well as theory predictions. Note that
a reliable reference for validating higher order bias is currently only provided by running  {\it separate universe}
N-body simulations \citep{Wagner2015, Lazeyras15}, which is beyond the scope of this article.

Alternatively to growth-independent bias measurements from third-order statistics one can employ
bias predictions from the peak-background split model for breaking the  growth-bias degeneracy.
In \citet[][hereafter referred to \paperIII]{HBG15-2} we tested linear bias predictions in the MICE-GC simulations,
confirming reports on inaccuracies of these predictions in the literature on a larger mass range thanks to
the large volume and resolution of the MICE-GC simulation. As in case of the third-order statistics no reliable
measurements for validating the predicted non-linear bias parameters were available. However, an 
interesting outcome of this analysis was a simple universal relation between the linear 
and non-linear bias parameters in the peak-background split model, which is independent of redshift and
cosmology for halo samples with $b_1 \gtrsim 2$.

In the present work we conclude our series of articles, presenting
results from an other method for allowing to measure bias parameters
in the MICE-GC simulation. This method is based on a direct
comparison of the halo and matter density fluctuations and may therefore be seen as the most direct way
of measuring bias parameters. We also study a variant of this method, which is based on
abundance matching
of tracers and matter density contrasts. The interest of this alternative method is that it can be applied in
galaxy surveys to estimate to bias function. Both methods deliver linear, as well as non-linear bias measurements.
We compare these new measurements with the most reliable results from our previous work (paper I 
and II, see Table \ref{table:references}), which allows for the validation of how well bias can be measured with each
approach.

The different measurements of linear and non-linear bias parameters furthermore allow for a
validation of the universal relation between the bias parameters, which we found for peak-background split 
predictions in \paperIII. \rev{The strength of our comparison emerges from the large halo mass range of
roughly $10^{12}-10^{15}$ \Msun, probed by the MICE-GC simulation as well as the fact that we use the same
halo mass samples throughout the whole comparison project.}

The remainder of this article is organised as follows. In Section \ref{sec:simulation} we describe the MICE-GC simulation 
as well as the halo samples on which our analysis is based on. The different methods for
obtaining the bias parameters are briefly reviewed in Section \ref{sec:bias_estimators}, while details on
bias measurements from the comparison of density contrasts are given in the appendix. We present our 
results in Section \ref{sec:results} which we summarise and discuss in Section 
\ref{sec:conclusion}.

\begin{table}
    \centering
      \caption{Abbreviations for previous articles of this series.}
      \label{table:references}
      \begin{tabular}{c  c c}
     name in the text & reference \\
      \hline
      \paperII & \citet{BHG15} \\
      \paperIII & \citet{HBG15-2}
           \end{tabular}
\end{table}

\section{Simulation and halo samples}\label{sec:simulation}

    Our analysis is based on the Grand Challenge run of the  Marenostrum
    Institut de Ci\`encies de l'Espai (MICE) simulation suite to which we refer to
    as MICE-GC in the following.
    Starting from small initial density fluctuations at redshift $z=100$ the formation of large scale
    cosmic structure was computed with $4096^3$ gravitationally interacting collisionless dark matter
    particles in a $3072$ $h^{-1}$Mpc box using the GADGET - 2 code \citep{springel05} with a softening
    length of $50$ $h^{-1}$kpc. The initial conditions were generated using the Zel'dovich
    approximation and a CAMB power spectrum with the power law index of $n_s = 0.95$, which
    was normalised to be $\sigma_8 = 0.8$ at $z=0$.  The cosmic expansion is described by the
    $\Lambda$CDM model for a flat universe with a mass density of
    $\Omega_m$ = $\Omega_{dm} + \Omega_b = 0.25$. The density of the
    baryonic mass is set to $\Omega_b = 0.044$ and $\Omega_{dm}$ is the dark matter density.
    The dimensionless Hubble parameter is set to $h = 0.7$. More details and validation test on this
    simulation can be found in \citet{mice1}.
    
    Dark matter haloes were identified as Friends-of-Friends groups \citep{davis85} with a linking
    length of $0.2$ in units of the mean particle separation. These halo catalogs and the corresponding
    validation checks are presented in \citet{mice2}. In the present analysis we 
    use the dark matter field as well as haloes in the comoving outputs at  redshift $z = 0.0$ and $0.5$.            
    The haloes are thereby divided  into the four redshift independent
    mass samples M0 - M3 (defined in Table \ref{table:halo_masses}), which span a mass range from
    Milky Way like haloes up to haloes of massive galaxy clusters.
    \begin{table}
    \centering
      \caption{Halo mass samples used in this study. $N_p$ is the number of dark matter particles per
      halo, $N_{halo}$ is the number of haloes per sample in the comoving output at redshift $z=0.5$.}
      \label{table:halo_masses}
      \begin{tabular}{c  c c c }
     \hline
     sample  & mass range [$10^{12} h^{-1}M_{\odot}$]   &  $N_p$ & $N_{halo}$\\
         \hline 
           M0	&	0.58 - 2.32	&	20-80		&	122300728 \\
           M1	&	2.32 - 9.26	&	80-316		&	31765907 \\
           M2	&	9.26 - 100	&	316-3416   &    8505326 \\
           M3	&	$\ge$ 100	&	$\ge$3416	&	280837 \\
          \hline
       \end{tabular}
    \end{table}

\section{bias estimators}\label{sec:bias_estimators}
        In this section we summarise the different methods for measuring and predicting linear and non-linear
        bias parameters. An overview of the methods with the abbreviations and references for the
        corresponding bias is given in Table \ref{table:definitions}.

        \subsection{$\delta_h-\delta_m$ relation}\label{sec:dhdm}
                
	In simulations we can measure the density contrasts of haloes and matter	(hereafter referred to as $\delta_h$
	and $\delta_m$ respectively), which allows for a direct estimation of the bias parameters
	$b_i$ in equation (\ref{eq:biasmodel_local}). In this sub-section, we describe the two methods which we employ
	for these measurements.
	%
	
\begin{figure}
             \includegraphics[width=80mm,angle=0]{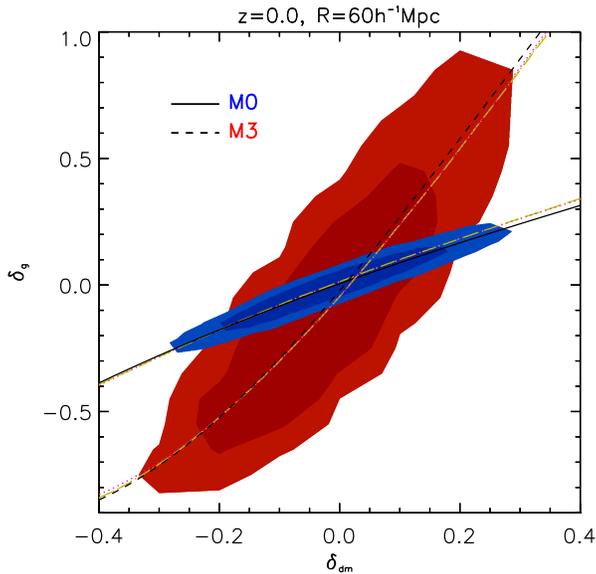}
          	\caption{\small  Scatter plot between halo and matter density contrast at $z=0$ obtained with a smoothing radius $R=60h^{-1}$Mpc. 
            	We represent the $68$\% (dark colour) and $95$\% (light colour) contour levels, in red we show the contour levels
          	corresponding to the mass bin M$3$ while in blue we show the mass bin M$0$.  
          	Black dashed lines show the bias function derived from the $\delta_h-\delta_m$ method, while the green long dashed,
		red dotted and magenta dotted lines show results from variants of the $\delta PDF$ method (see text for details).
             }
            \label{fig:dg-dm_r60}
     \end{figure}

	\rev{The first method has been widely used in the literature and may be seen as the 
	most direct way of measuring bias, provided that the biasing function is deterministic and local.
	It consists in fitting a polynomial of order $n$ to the $\delta_h(\bf r_i) -\delta_m(\bf r_i)$
	relation, which can be measured from the smoothed density fields of haloes and matter
	at different positions $\bf r_i$ in a simulation.
	The scatter between haloes and matter density contrasts in the  MICE-GC simulation is shown
	in Fig.~\ref{fig:dg-dm_r60} at redshift $z=0.0$ for a smoothing scale of $R=60$\Mpc . One can see that
	the average relation between $\delta_m$ and $\delta_h$ is different regarding the halo sample: the slope
	of the most massive one (M3) is higher than the less massive (M0). This is expected because the linear bias
	which controls the slope of the $\delta_h$-$\delta_m$ relation is higher for the sample M3.
	In general, the best-fit parameters of the polynomial of order
        $n$ fitted to the scatter plot are then identified as the bias coefficients.
         Note that the parameter $b_0$ is in principle constrained by requiring that by definition the average of the
         halo density contrast is null, $\langle \delta_h \rangle = 0$. In case of a cubic biasing relation it leads to the formal expression}

        \begin{equation}
        b_0=-b_1\sigma_{dm}^2\left\{ \frac{c_2}{2} + \frac{c_3}{6}S_{3,dm}\sigma_{dm}^2 \right \},
        \label{eq:b0}
        \end{equation}
        \rev{where $S_{3,dm}$ and  $\sigma_{dm}^2$ are respectively the skewness  and the variance of the dark matter field smoothed
        on scale $R$. Note that $c_3$ is the cubic bias coefficient defined as $c_3\equiv b_3/b_1$. We will refer to this method as
        $\delta_h-\delta_m$ in the remainder of this article.}

        \rev{In practice, the choice of the order of the polynomial could affect the estimation of the bias coefficients 
	$b_1$, $c_2$ and $c_3$ \citep[see][]{Lazeyras15}. One can expect that this truncation introduces a dependence of 
	the bias parameters on the smoothing scale, used to estimate the density contrasts. We therefore perform the same
	convergence test proposed by \citet{Lazeyras15}, which consists in comparing the bias coefficients obtained with a third-,
	fourth- and fifth-order polynomial. In the case of third-order polynomials we test, in addition, the impact of the choice of $b_0$ on
	the estimated bias parameters. 	To do so we let $b_0$ being free and compare the resulting bias parameters to those derived
	fixing $b_0$ to the expected value from equation (\ref{eq:b0}). Note that, we expect the local deterministic bias to describe
	the mapping between matter and haloes only on sufficiently large smoothing scales.}
	
\begin{figure}
           \includegraphics[width=80mm,angle=0]{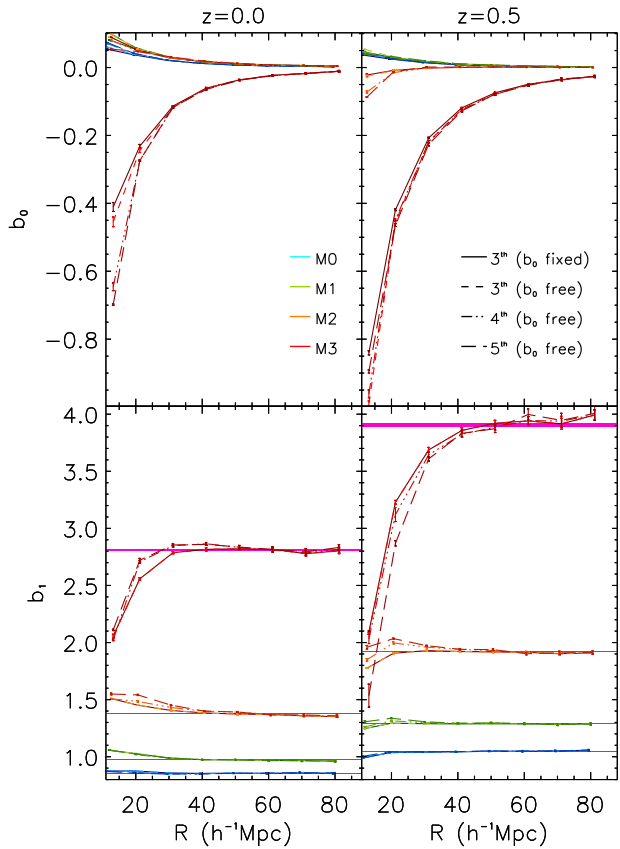}
    	\caption{\small  {\it Top:}  Bias coefficient $b_0$. {\it Bottom:} Linear bias coefficient $b_1$.
    	In both panels, we estimate quantities with respect to the smoothing scale $R$. We test several methods:
    	in the first we fix $b_0$ (solid lines) using equation (\ref{eq:b0}), then we allow $b_0$
    	to vary as a free parameter (short dashed lines), after we fit a fourth (dashed dot-dot-dot lines) and finally a
    	fifth (long dashed lines) order polynomials.  In the bottom
        panels we also represented the results of the fit (magenta) of
    	the linear bias $b_1$ between $40$ and $80h^{-1}$Mpc. We perform this analysis for two comoving
    	outputs at redshift $0.0$ (left) and $0.5$ (right). In all panels we adopt a colour coding blue, green, orange and
    	red referring respectively to the halo samples M$0$, M$1$, M$2$ and M$3$. Note that, for clarity, each method
    	applied to a given halo mass is represented with a different shade of the corresponding colour.}
    	\label{fig:results8}
    \end{figure}

\begin{figure}
            \includegraphics[width=80mm,angle=0]{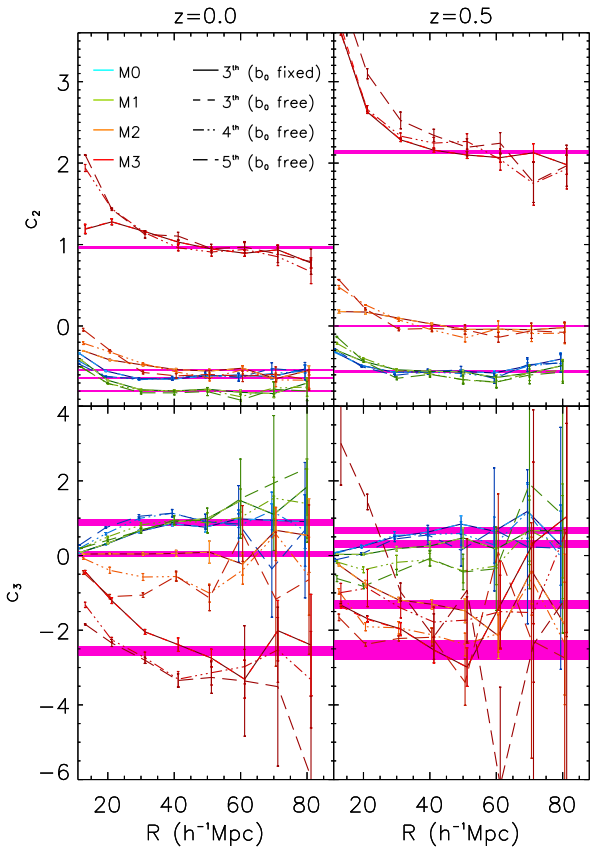}
    	\caption{\small {\it Top:} Second-order bias coefficient $c_2$. {\it Bottom:} Third-order bias coefficient $c_3$.
    	We adopt the same colour and line coding described in Fig. \ref{fig:results8}}
    	\label{fig:results9}
    \end{figure}

    \rev{The outcome of this analysis is presented is Fig. \ref{fig:results8} and \ref{fig:results9}, which show the results of this fitting process for
    respectively $b_0$, $b_1$ and $c_2$, $c_3$ when applied at various smoothing scale $R$.
    For each scale, we estimate the error by performing the fit in $64$ Jackknife sub-samples of the full
    simulation. Examination of the top panel of Fig.~\ref{fig:results8} shows that $b_0$ 
    changes with scale, while converging to zero. Furthermore its sign 
    depends on the mass of the halo sample. Both effect are expected, as shown by equation (\ref{eq:b0}), the sign of
    $b_0$ can change, depending on the higher order bias
    coefficients. Moreover, since $b_0$ depends on quantities
    such as the variance $\sigma^2$ and the skewness $S_3$ of matter,
    we expect a scale dependence. 
    Regarding the biasing coefficients, Figures \ref{fig:results8} and \ref{fig:results9} show that we find a strong scale dependence of our
    bias measurements for small smoothing radii $R$, while results converge for scales larger than $R = 40$ \Mpc\ \citep[see also][]{M&G11}.
    In general, we can also conclude that the choice of leaving $b_0$ free or not does not impact the fitted bias parameters. In addition,
    looking at the two low mass samples we can see that above $40h^{-1}$Mpc, the choice of the order of the fitted polynomial does not
    impact significantly the measurements. Despite, the visual difficulty of interpreting the results for the two high mass samples we will
    show later that this choice can have an impact on the fitted third-order bias parameter. We therefore only use scales between
    $40 \le R  \le 80$ \Mpc\ to derive the bias parameters via the $\delta_h-\delta_m$ method.}

    %

       \rev{The second method for measuring bias parameters from the $\delta_h-\delta_m$ relation consists in matching
       the probability distribution functions (PDF) of $\delta_h$ and $\delta_m$, it will therefore be referred to as
       the $\delta\text{PDF}$ method in the following. As the $\delta_h-\delta_m$ method, it relies on the
       assumption that the bias function is local and deterministic. 
       However, the $\delta\text{PDF}$ method presents a clear advantage: it can be directly applied to measure galaxy bias from
       observations. The PDF of matter fluctuations can be, indeed, modelled for a given cosmology and the PDF of galaxies
       can be directly observed from redshift surveys
       \citep[see][]{bel2016,DiPorto2015, mar05, bernardeau02, Uhlemann16}. Inferring the bias from the
       $\delta_h-\delta_m$ method, on the other hand, requires adequate observations for measurements of
       $\delta_m$ via weak gravitational lensing and introduces additional uncertainties \citep{Chang16, Pujol16}.}
     
      \rev{Thus, measuring the bias parameters by matching the $1$-point probability distribution of haloes and matter
      fluctuations ($P_h(\delta_h)$ and $P_m(\delta_m)$ respectively) represents an alternative way of estimating
      local bias parameters. It consists in assuming the existence of a local mapping between the matter and haloes
      fluctuations such that the probability is conserved under a change of variable following this functional relation}

    \begin{equation}
    \mathcal{C}_h[\delta_h] = \mathcal{C}_m[\delta_{m}],
    \label{eq:cumu}
    \end{equation}
    \rev{where $\mathcal{C}_i$ is the cumulative distribution function, defined as
    $\mathcal{C}_i[\delta_i]\equiv \int_{-1}^{\delta_i}P_i(\delta^\prime_i)d\delta^\prime_i$ 
    for either haloes ($i=h$) or matter ($i=m$).
    The halo density fluctations $\delta_h$ can thus be expressed as a function of
    halo density fluctuations $\delta_m$ by inverting equation (\ref{eq:cumu})
    (i.e. $\delta_h(\delta_m) = {\mathcal{C}_h}^{-1}\left [\mathcal{C}_m[\delta_{m}] \right ]$),
    which is by definition the local bias function. The exponent $-1$ denotes the reciprocal function. The mapping
    $C_h ^{-1}[C_m]$ is obtained by integrating numerically the probability distribution function of both the
    halo and matter density contrasts. For technical reasons, we use a cubical smoothing of size $64$\Mpc\ which corresponds in
    volume to a spherical smoothing of radius $R\simeq 40$\Mpc\  (which, we saw, is large enough to consider the bias coefficients
    as scale independent). Note that when the average number of haloes per cell is smaller than $150$ (which is the case for
    the samples M2 and M3), it is necessary to correct from shot noise effects \citep[see][]{DiPorto2015}. 
    For reconstructing the halo PDF we use the relation between the discrete probability
    distribution $P_N$ (where $N$ represents the number of haloes inside a cell) and the probability
    density function of halo density fluctuations $P_h(\delta_h)$,}
    
    \begin{equation}
    P_N=\int_{-1}^\infty K[N|\bar N(1+\delta_h)]P_h(\delta_h)\dif\delta_h,
    \label{eq:samplingd}
    \end{equation} 
    \rev{where $K$ is the conditional sampling probability which we assume to be a Poisson distribution \citep{layser}.
    In practice the inversion of equation (\ref{eq:samplingd}) requires some subtleties \citep[see][]{bel2016}. Following \citet{bel2016}, we apply
    three different reconstruction methods and choose the one which provides the best description of the counting
    probability distribution $P_N$. Given the high shot noise level in the $M2$ and $M3$ samples we assume
    a parametric form for the density probability $P_h$, in practice we tested a skewed-LogNormal \citep{colombi94}
    and a Gamma expansion \citep{GF&E2000} and found that the Gamma expansion offers a better matching of the
    measured counting probability of haloes.}
    \begin{figure}
    	\includegraphics[width=90mm,angle=0]{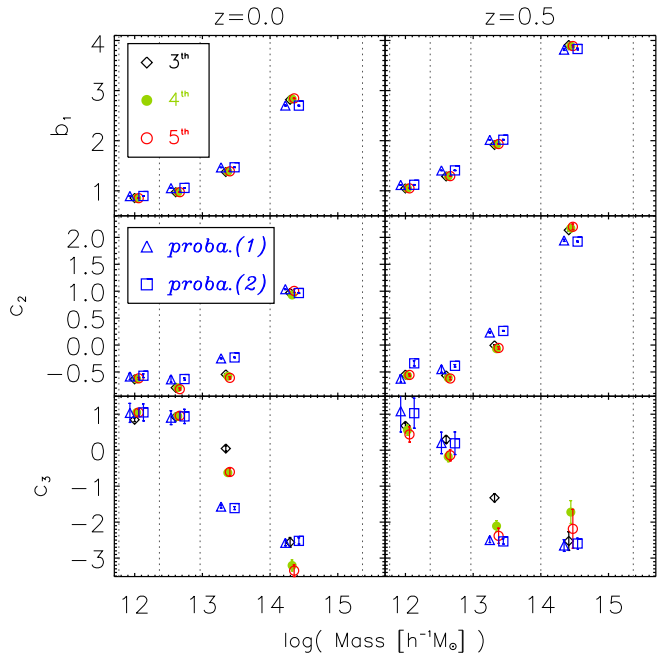}
    	\caption{\small  On one hand, we display the bias coefficients ($b_1$, $c_2$ and $c_3$ from top to bottom) obtained
    	by fitting the scatter plot with increasing order of the polynomial (black diamonds, green filled circles and orange empty
    	circles) on scales between $40$ and $80h^{-1}$Mpc with respect to the average mass of the halo samples. On the
    	other hand, we show the obtained bias coefficients from the two variants of the $\delta\text{PDF}$ method (see text for more details). } 
    	\label{fig:sp}
    \end{figure}  
    
    \rev{Once we obtained the bias function we devised two ways of estimating the bias parameters. In both cases
    we fix an effective range on which the Taylor expansion must be valid and on one hand we fit the bias
    function with a third-order polynomial (referred to as
    \textit{proba. (1)}) and on the other hand we apply discreet derivatives
    method (referred to as \textit{proba. (2)}). We
    allow the effective range in $\delta$ to decrease and stop when the two methods converge to the same
    values for $b_1$, $c_2$ and $c_3$. An illustration of this technique is displayed in Fig. \ref{fig:dg-dm_r60},
    where we show in green long dashed line the bias function obtained from inversion of
    equation (\ref{eq:cumu}) together with the Taylor expansion estimated from the fitting method
    (orange dotted lines) and from the numerical derivatives method (magenta dotted lines).
     One can see that the results from the two methods agree well with the obtained bias function from the probability
    density function. On the other hand, we can see some significant departures from the Taylor expansion
    obtained from the direct fit to the scatter plot (black lines). In fact, considering Fig. \ref{fig:sp} which
    summarizes the measurements of the bias coefficients from the two presented methods one can remark
    that despite small discrepancies the overall behaviour of the bias coefficients with respect to the mass is
    qualitatively in good agreement. In Fig. \ref{fig:sp} we also show, with more clarity, the impact of the choice
    of the order of the fitted polynomial to the scatter plot: it becomes clear that in the halo samples M2 and M3
    there is a non-negligible (given the estimated error bars) effect on the estimation of the $c_3$ coefficient
    while the lowest order bias coefficients are not significantly affected.}

    \subsection{two-point cross-correlation}\label{sec:bxi}
    
    The bias function in equation (\ref{eq:biasmodel_quadratic_nonlocal}) has an important application in the cosmological
    analysis of galaxy clustering statistics, since it provides a parameterisation 
    of the deviation between the clustering of haloes (or galaxies) and the underlying 
    matter distribution. In simulations, where the clustering of matter can be directly measured,
    one can infer the bias parameters by comparing the clustering statistics of haloes and matter.
    In this subsection we focus on bias measurements from the two-point correlation $\xi$ as statistical
    measure for the clustering. It directly probes the linear growth factor, thereby 
    providing strong constraints on cosmological models.
    The two-point correlation can be defined in configuration space as the mean product of density fluctuations
    $\delta_i$ at the positions $\bf{r_1}$ and $\bf{r_2}$ that are separated by the distance
    $r_{12} \equiv \vert \boldsymbol{r_1}-\boldsymbol{r_2} \vert$,
    \begin{equation}
    \xi_{xy}(r_{12}) \equiv
    \langle  \delta_x(\boldsymbol{r_1}) \delta_y(\boldsymbol{r_2}) \rangle =
    \langle \delta_x^1 \delta_y^2 \rangle(r_{12}).
    \label{eq:def_2pc}
    \end{equation}
    The indices $x$ and $y$ refer to the density fields of haloes (or galaxies) 
    and matter, while $\langle \dots \rangle$ denotes the average over pairs of any 
    orientation and position. The two-point auto-correlation (hereafter referred to as 2pc)
    corresponds to the case $x=y$ and will be denoted in the following as $\xi_h \equiv \xi_{hh}$  and
    $\xi_{m} \equiv \xi_{mm}$ for haloes and matter respectively,
    whereas the  halo-matter cross-correlation (hereafter referred to as 2pcc) 
    will be denoted as $\xi^{\times} \equiv \xi_{hm}$.
    Inserting the local bias model from equation (\ref{eq:biasmodel_quadratic_nonlocal}) into equation (\ref{eq:def_2pc})
    provides, at leading order, a simple relation between $\xi_{m}$ and 
    $\xi^{\times}$,
    
    \begin{equation}
    \xi^{\times}(r_{12})  \simeq ~b_1~\xi_{m}(r_{12}) + {\cal O}[\xi_{m}^2].
    \end{equation}
    At large scales ($r_{12}\gtrsim20 h^{-1}$Mpc), where the two-point correlation function is small,
    we expect ${\cal O}[\xi_{m}^2]$ to be  negligible (see e.g. \paperII), which allows for measurements of the
    linear bias as
    	
    \begin{equation}
    b_{\xi}^{\times}(r_{12}) \equiv\sqrt{ \frac{\xi^{\times}(r_{12})}{\xi_{m}(r_{12})}}
    \simeq b_1.
    \label{eq:bxicross}
    \end{equation}
    We showed in \paperII\ that $b_{\xi}^{\times}$ is a reliable estimate of the linear bias 
    $b_1$, since it agrees with various other estimators from second-order one- and two-point
    statistics at large scales.
    
    \subsection{three-point auto- and cross-correlations}\label{sec:dQ}
    Third-order statistics is sensitive to the shapes of large-scale density fluctuations. It therefore
    provides information about the large-scale structure which is not accessible with second-order one
    or two-point statistics as the latter is defined isotropically in configurations space.
    Combining second- and third-order statistics therefore allows for breaking the degeneracy between
    the linear growth factor and the linear bias parameter which otherwise limits the accuracy of 
    cosmological constraints derived from the 2pc at large scales \citep[see e.g.][or \paperII]{bernardeau02}.
    However, this approach relies on the accuracy with which the linear bias can be measured with
    third-order statistics.
    The most general third-order statistics in configuration space is the three-point correlation or its reduced 
    counter part \citep{grope77}. In this subsection we describe our bias measurements
    from \paperII, which are derived from the 
    symmetrised reduced three-point halo-matter-matter cross-correlation in configuration space
    \citep[hereafter referred to as 3pcc, see also][]{Pollack12}. It is defined as
    \begin{equation}
    Q^{\times}\equiv \frac{1}{3}\frac{\zeta^{hmm}  + \zeta^{mhm} + \zeta^{mmh}}{\zeta^{hm}_H},
    \label{eq:def_Q3cross}
    \end{equation}
    where
    \begin{equation}
    	\zeta^{xyz}  \equiv \langle  \delta_{x,1} \delta_{y,2} \delta_{z,3} \rangle
    	\label{eq:def_3pc}
    \end{equation}	
    is the three-point halo-matter-matter cross-correlation and
      	
    \begin{equation}
    	\zeta^{hm} _H \equiv \xi^{hm}_{12} \xi^{hm}_{13}  + \xi^{mh}_{12} \xi^{hm}_{23}  + \xi^{mh}_{13}
    	\xi^{mh}_{23}
    	\label{eq:def_3pcH}
    \end{equation}
    is the corresponding hierarchical three-point cross-correlation.
    The average $\langle \dots \rangle$ is made over triangles of any 
    orientation and position. The auto-correlation for halo and
    matter density fields ($Q_h$ and $Q_m$ respectively) are defined analogously and will be referred to as 3pc in the following.
    For measuring the bias accurately using the 3pcc one needs to take into account non-local contributions to
    the bias function. Neglecting the latter would cause a $\sim20\%$ error 
    in the estimation of the linear bias parameter \citep[see ][]{M&G11, 
    chan12,baldauf12}. We therefore proposed in \paperII\ to measure the 
    linear bias independently of quadratic local and non-local contributions by combining three-point auto- and
    cross-correlations,
    \begin{equation}
    	b_{\Delta Q} \equiv - 2 \frac{Q_m}{Q_h-3 Q^{\times}_h}.
	\label{eq:bdQ}
\end{equation}
    A similar combination of auto- and cross-correlations allows for
    the determination of the quadratic local and non-local bias parameters ($c_2$ and $\gamma_2$)
    independently of $Q_m$, once the linear bias is determined (for instance via equation 
    (\ref{eq:bdQ}) or the 2pc),
    \begin{equation}
        \Delta Q_{cg} \equiv Q_h-Q^{\times}=
        \frac{2}{3}\frac{1}{b_1}
        \left[c_2 + g_2 Q_{nloc}\right ].
        \label{eq:cdQ}	
    \end{equation}
	We predict the non-local contribution $Q_{nloc}$ from the non-linear power spectrum,
	which has been measured in the simulation. Details on this prediction are provided in the appendix of
	\paperII.
	Combining three-point auto- and cross-correlation for measuring bias will be 
	referred to as $\Delta Q$ method in the following. The
    results in this work are derived from 3pc measurements which are computed using triangles with
    fixed legs of $36$ and $72h^{-1}$Mpc at $18$ opening angles between 
    these legs. The density fields where smoothed with an $8$\Mpc\ Top-hat window function,
    and the covariance between measurements at different opening angles was estimated from $64$ JK
    samples. Details on these measurements are presented in \paperII, where we also
    study the dependence of 3pc bias measurements on the triangle scale.
        
	\subsection{third-order cross-moments}\label{sec:tau}

       Alternatively to the 3pc one can explore the third-order statistical properties of cosmic fields
       with one- and two-point statistics. Despite the fact that they do not provide access to the full
	third-order hierarchy probed by the 3pcc (Subsection \ref{sec:dQ}) they are useful tools 
	for comparing the statistical properties of the halo distribution with respect to the underlying matter field. 
	
	These one- and two-point statistics correspond to the 3pc for very specific triangle 
	configurations. The 3pc for triangles which are collapsed into two points corresponds to the
	correlator $C_{12}$ \citep[see][]{bernardeau96}, while for a
        further collapse into a single point yields the skewness $S_3$ \citep[see][]{bernardeau02}.
	As in case of the 3pcc, it is possible to define the cross-skewness $S_{3}^\times$ and the
	cross-correlator $C_{12}^\times$ as
	
      \begin{equation}
      S_{3}^\times \equiv \frac{\langle\delta_h\delta_m^2\rangle}{\langle\delta_h\delta_m\rangle^2}
      \label{eq:S3rosss}
      \end{equation}     
       and
     
      \begin{equation}
       C_{12}^\times \equiv \frac{\langle\delta_{h,1}\delta_{h,2}\delta_{m,2}\rangle}{\langle\delta_{h,1}\delta_{h,2}\rangle\langle\delta_h\delta_m\rangle}
      \label{eq:C12rosss}
      \end{equation}
	where $\delta_h$ and $\delta_m$ are respectively the density contrast of halo and matter density fields
	which are smoothed with a spherical top-hat window function of radius $R$. The auto skewness
	$S_{3}$ and the auto-correlator $C_{12}$ of the matter density field are defined analogously. We provide
	details on how we measure these quantities and how to correct them for shot-noise in \paperII.
       	One can express the linear and quadratic bias parameters independently from each other using the combinations
	
	\begin{equation}
  	       b_{\tau}^{\times}\!\!\!  \equiv  \!\!\frac{{\displaystyle \,S_{{3}}-\,C_{{12}} }}{{\displaystyle \,S_{{3}}^\times -\,C_{{12}}^\times }}\equiv \frac{\tau}{\tau^\times}
        	\label{eq:btaucross}
	\end{equation}
       and
	\begin{equation}
        	c_{\tau}^{\times}\!\!\! \equiv \!\!\frac{{\displaystyle S_{{3}}C_{{12}}^\times- C_{{12}}S_{{3}}^\times }}{{\displaystyle  \tau^\times }}.
	\label{eq:ctaucross}
	\end{equation}
	We will refer to this way of measuring bias as the $\tau^\times$-method in the 
	following.  For spherically averaged quantities such as skewness, correlators,
	cross-skewness and cross-correlators non-local contributions of the second-order described by 
	equation (\ref{eq:G2}) lead to an effective second-order local bias
      
      \begin{equation}
      c_2^{\mathrm{eff}}=c_2 - \frac{4}{3}\frac{\gamma_2}{b_1},
      \label{eq:c2eff}
      \end{equation}
      but do not affect the estimated linear bias parameter $b_1$ unless for 
      highly biased samples (M3, see discussion in \paperII). The same effect has 
      been pointed out by \citet{chan12} for the $\delta_g-\delta_m$ method for measuring 
      bias, which is described in Section \ref{sec:dhdm}.
      Assuming a local Lagrangian bias model, $\gamma_2 = -\frac{2}{7}(b_1-1)$ 
      we expect the non-local contributions to $c_2^{\mathrm{eff}}$ to be zero 
      for $b_1=0$ and $\lesssim 0.3$ for $b_1=4$. However these expectation 
      will not be accurate as moderate deviations from the Lagrangian bias model
      have been reported by \citet{chan12} and in \paperII.

	\subsection{Peak-background split predictions}\label{sec:pbs}

      The peak-background split (hereafter referred to as PBS) model provides
      predictions for the linear and non-linear bias parameters as a function of halo 
      mass, which are deduced from derivatives of the mass function.
      Model fits to mass function measurements therefore provide analytical expression of the bias as a function
      of halo mass.
      Such PBS bias predictions are essential for various purposes like
      bias modelling  as function of galaxy properties via Halo Occupation 
      Distribution (HOD) modelling, cluster mass calibration
      or predictions of the cluster count and power spectra covariances
      \citep[e.g.][]{CooraySheth02, LimaHu04, LimaHu05, Lacasa16}.
      It is therefore important to verify how
      accurately the PBS model is able to predict bias measurements in simulations.

      In the present analysis we validate the accuracy of PBS predictions for
      the linear, quadratic and  third-order bias parameters by comparing them to the various 
      measurements described in the previous subsections.
      \rev{We employ fits to the measured MICE-GC mass function from \paperIII, based on 
      the models of \citet{Tinker10} and \citet{warren06} as well as a new model presented in \paperIII, to which we
      refer to as Tinker, Warren and HBG15 model respectively in the following. The sample M0 has been
      excluded from the fitting range since we expect halo detection in the corresponding mass range to be very noisy,
      as we discuss in \paperIII. However, the predictions can still be made over the full mass range for all the samples M0-M3.
      For studying the universality of the relation between the different bias parameters we adopt the fitting parameters 
      for the Tinker and Warren model, which are provided in the corresponding articles. Note that these latter fits are
      based on simulations with different cosmologies than in the MICE-GC simulation. Furthermore \citet{Tinker10}
      define haloes as spherical over-densities, not as FoF groups as done by \citet{warren06} and in the present study.}
      
\section{Results}\label{sec:results}

\subsection{bias comparison}

We compare in Fig. \ref{fig:bias_comparison} the linear and non-linear bias parameters from the different
measurements and the PBS predictions described in Section \ref{sec:bias_estimators} using the
mass samples M0-M3 (defined in Table \ref{table:halo_masses}) at the redshifts $z=0.0$ and $z=0.5$.

\begin{table*}
      \centering
        \caption{Bias estimators, definitions and notations. The term cross refers to halo-matter cross
        correlations, $b$ and $c$ indicate linear and non-linear bias estimations respectively.}
        \label{table:definitions}
        \begin{tabular}{c c c c }
        & method & bias parameters  &   reference\\
           \hline 
             measurements& $\delta_h - \delta_m$  & $b_1, c_2, c_3$  & Section \ref{sec:dhdm}  \\
            & $\delta$PDF & $b_1, c_2, c_3$  & Section \ref{sec:dhdm}  \\
             &two-point cross-correlation $\xi^\times$	  &	$b_1$ & Section\ref{sec:bxi}, \paperII  \\
             &combined three-point auto- and cross-correlations $\Delta Q$ & $b_1, c_2$ & Section\ref{sec:dQ}, \paperII \\
             &combined third-order auto- and cross-moments $\tau^\times$  & $b_1, c_2$  & Section\ref{sec:tau}, \paperII \\
              \hline
             predictions & peak-background split (PBS) model &  $b_1, c_2, c_3$ & Section\ref{sec:pbs}, \paperIII \\
         \end{tabular}
      \end{table*}
      
\subsubsection{linear bias}

In the case of the linear bias ($b_1$, top panel of Fig. \ref{fig:bias_comparison}) we find an overall variation
of roughly $5$ percent between results from different measurements and PBS predictions 
at all masses and both redshifts.
%
The measurements from the 2pcc (\bxicross, equation (\ref{eq:bxicross})) are expected to be the most
reliable estimate with percent level accuracy (see e.g. \paperII ). We therefore use it as a reference 
for evaluating the accuracy of the other linear bias measurements and predictions in our 
comparison.


The results in Fig. \ref{fig:bias_comparison} show that both estimations for the linear bias
from third-order statistics,
($\Delta Q$ and $\tau^\times$, described in Section \ref{sec:dQ} and \ref{sec:tau} respectively)
are in $1\sigma$ agreement with \bxicross. However, the errors of \btaucross are roughly  $10$ percent, while those of
\bdeltaq at the one percent level. We attributed the strong deviation of \btaucross from all 
other results at high masses to an inadequate shot-noise correction in \paperII. From the same
analysis we expect larger deviations from \bxicross
when the linear bias is measured from third-order auto- instead of cross-correlations correlations
and when non-local contributions are not taken into account.

The two types of linear bias measurements from the $\delta_h-\delta_m$ relation
(described in Section \ref{sec:dhdm}) are both in percent level agreement with \bxicross
which is consistent with results from \citet{M&G11} or \citet{Pollack12}.
Results from the $\delta$PDF method tend to be overall slightly larger with $\lesssim7\%$ deviations
from \bxicross. \rev{A possible reason for the discrepancies can be inaccuracies in the covariance estimation 
from Jackknife sampling, which might affect the best fit values. Note that we do not expect the discrepancies
to result from the truncation of the  bias function at third-order based on our convergence tests in Section
\ref{sec:dhdm}.}

In addition to the bias measurements we show predictions from the PBS model
which are based on different fits to the MICE-GC mass function, presented in \paperIII. The different 
predictions show a very good mutual agreement in the mass range spanned by the samples 
M1-M3 which was used for the fits. For the low mass  sample, which was excluded
from the fitting range due to noisy halo detection, the  different PBS predictions show a stronger variation.
For the mass samples M2 and M3 ($\gtrsim 10^{13}$\Msun) all predictions lie up to $\sim 7$ percent
below \bxicross, which is consistent with findings in the literature \rev{\citep[e.g.][]{MS&S10, M&G11, Pollack12,
Paranjape13b, Lazeyras15}.}

The agreement between predictions from different mass function fits indicates that the inaccuracy of PBS bias
is driven by shortcomings of the model, rather than uncertainties in the fits.
\rev{
Such a shortcoming could be the assumption of a constant matter
density threshold for the gravitational collapse
on which the standard PBS model application to mass functions is based on, as pointed out in recent studies.
In particular \citet{Paranjape13b} showed that the scale-independent PBS bias parameters, reconstructed from scale dependent
Lagrangian bias measurements in simulations, are in good agreement with predictions from the excursion set peak
\citep[ESP,][]{Paranjape13a} model, when a halo mass-dependent scatter in the collapse threshold is included.
A percent level agreement between such ESP predictions for PBS bias and direct measurements of the latter from separate
universe simulations (which do not rely on a threshold model) was reported by \citet{Lazeyras15}.
These direct PBS bias measurements are also in excellent agreement with bias measurements from large-scale Fourier
space clustering, confirming results from \citet{Baldauf15} and \citet{Li15}.}

\subsubsection{non-linear bias}

\begin{figure*}
	\includegraphics[width=175mm,angle=270]{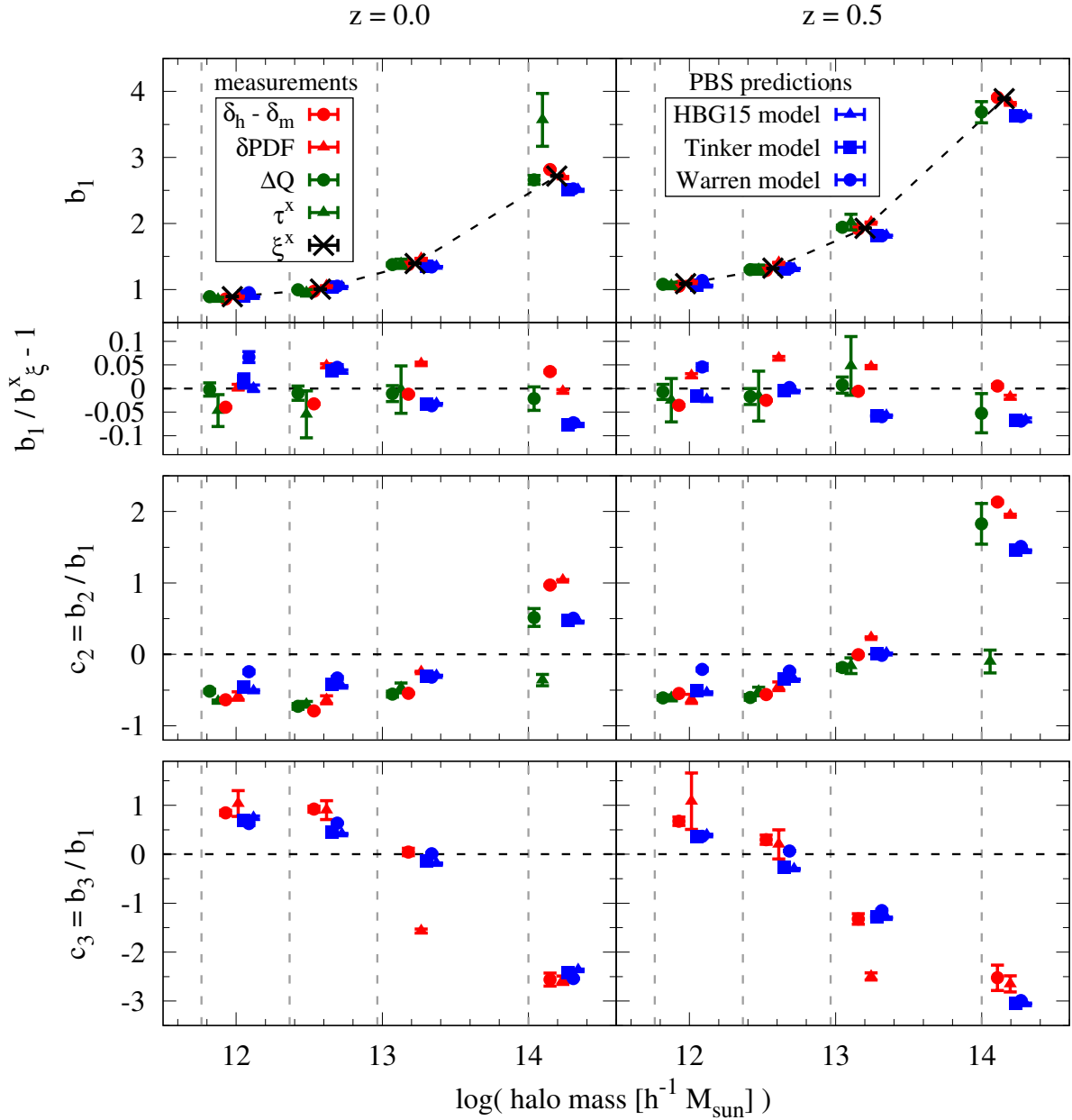}
	\caption{
	Summary of various bias measurements from second- and third-order halo-mater 
	cross-correlations ($\xi^{\times}$, $\Delta Q$ and $\tau^{\times}$), $\delta_h-\delta_m$ 
	relations	and PBS predictions, described in Section \ref{sec:bias_estimators}. The top, central and bottom
	panel show the first- second- and third-order bias parameters respectively at $z = 0.0$ and $0.5$
	(left and right panel) versus the mean halo mass of each mass samples M0-M3
	(slightly shifted along the mass axis for clarity). The subpanel at the bottom of the top panel shows relative deviations
	from $\xi^{\times}$.
	The lower and upper limits of the mass samples are marked by vertical
	grey dashed lines. Error bars denote $\sigma$ uncertainties.}
	\label{fig:bias_comparison}
\end{figure*}

Measurements and prediction of the second- and third-order bias parameters ($c_2 \equiv b_2/b_1$ and
$c_3 \equiv b_3/b_1$) are shown in the central and bottom panels of Fig. \ref{fig:bias_comparison} respectively. 
Note that for this comparison we do not have a reference for the non-linear bias, \rev{such as
measurements from separate universe simulations.}
Furthermore our measurements from third-order statistics provide only bias parameters up to
second-order, since they are based on leading order perturbation theory. Measurements of $c_3$ are  therefore
only obtained from the $\delta_h-\delta_m$ relation.

Overall we find a stronger variation between the different results for the 
non-linear bias than for the linear bias. Yet, all the different non-linear bias
measurements and predictions show similar mass dependencies at both redshifts.
The second-order bias $c_2$ depends only weakly on the halo mass below
$\lesssim 10^{13} h^{-1}M_{\odot}$ (covered by the mass samples M0 and M1), with a
value of $\simeq -0.5$ at both redshifts. In the mass range
$\simeq 10^{13}-10^{14} h^{-1}M_{\odot}$, which corresponds to the mass sample M2,
$c_2$ changes from negative to positive values, and increases rapidly with mass to values of
up to $2$ at $z=0.5$ for halo masses of higher than $10^{14} h^{-1}M_{\odot}$.
The PBS prediction for $c_2$ from all mass function fits show an overall weaker mass dependence than the 
corresponding measurements, as the former lie above the results from the $\Delta Q$, $\tau^\times$
and $\delta_h-\delta_m$ methods in the mass range spanned by M0-M2, where $c_2$ is 
negative. For the high mass sample M3, where $c_2$ is positive at both redshifts we find the 
opposite. \rev{The latter result confirms findings of \citet{M&G11} and \citet{Lazeyras15} on a larger mass range.
An overly weak mass dependence of standard PBS predictions has also been found for Lagrangian bias by \citet{Paranjape13b}.}

In the mass range M0-M2 the measurements  from the $\Delta Q$, $\tau^\times$ and $\delta_h-\delta_m$ methods agree mutually at
the $1\sigma$ level at both redshifts, while results vary strongly for the high mass sample 
M3. Note that we expect the $c_2$ measurements from the $\delta_h-\delta_m$ and 
$\tau^\times$ to be biased by non-local contributions with values between zero for $b_1=0$ and
$\lesssim 0.3$ for $b_1=4$, as discussed in Section \ref{sec:tau}, which is comparable with the variation
among the different results.

In the bottom panel of Fig. \ref{fig:bias_comparison} we present a validation of third-order
bias prediction from the PBS model with direct measurements from the $\delta_h-\delta_m$
and $\delta$PDF methods.
The predictions and measurements are in an overall agreement with each other, as both decrease
from positive values of around unity for M0 to negative values of down to $\lesssim -3$
for M3 at both redshifts. The zero crossings of $c_3$ occur between $10^{12.5} - 10^{13.5} M_{\odot}/h$,
\rev{which is consistent with the direct PBS bias measurements from separate universe simulations from \citet{Lazeyras15}. 
Overall the predictions tend to be smaller than the measured values. Note that we used
third-order polynomials for the bias function in order to measure the bias parameters. From our convergence test in Section
\ref{sec:dhdm}, we expect larger deviations from PBS predictions in the high mass bins M2 and M3 
when the third-order bias is measured using higher order polynomials.}

\subsection{Universal relation between bias parameters}

Our various bias measurements enable us to validate the universal relation between the 
linear and non-linear bias parameters $b_2(b_1)$ and $b_3(b_1)$, which we found 
for PBS predictions and linear bias values of $b_1 \gtrsim 2$ in \paperIII. Such a universal
behaviour can be expected from a universality of the mass function for different redshifts and
cosmologies.

\begin{figure*}
	\includegraphics[width=105mm,angle=270]{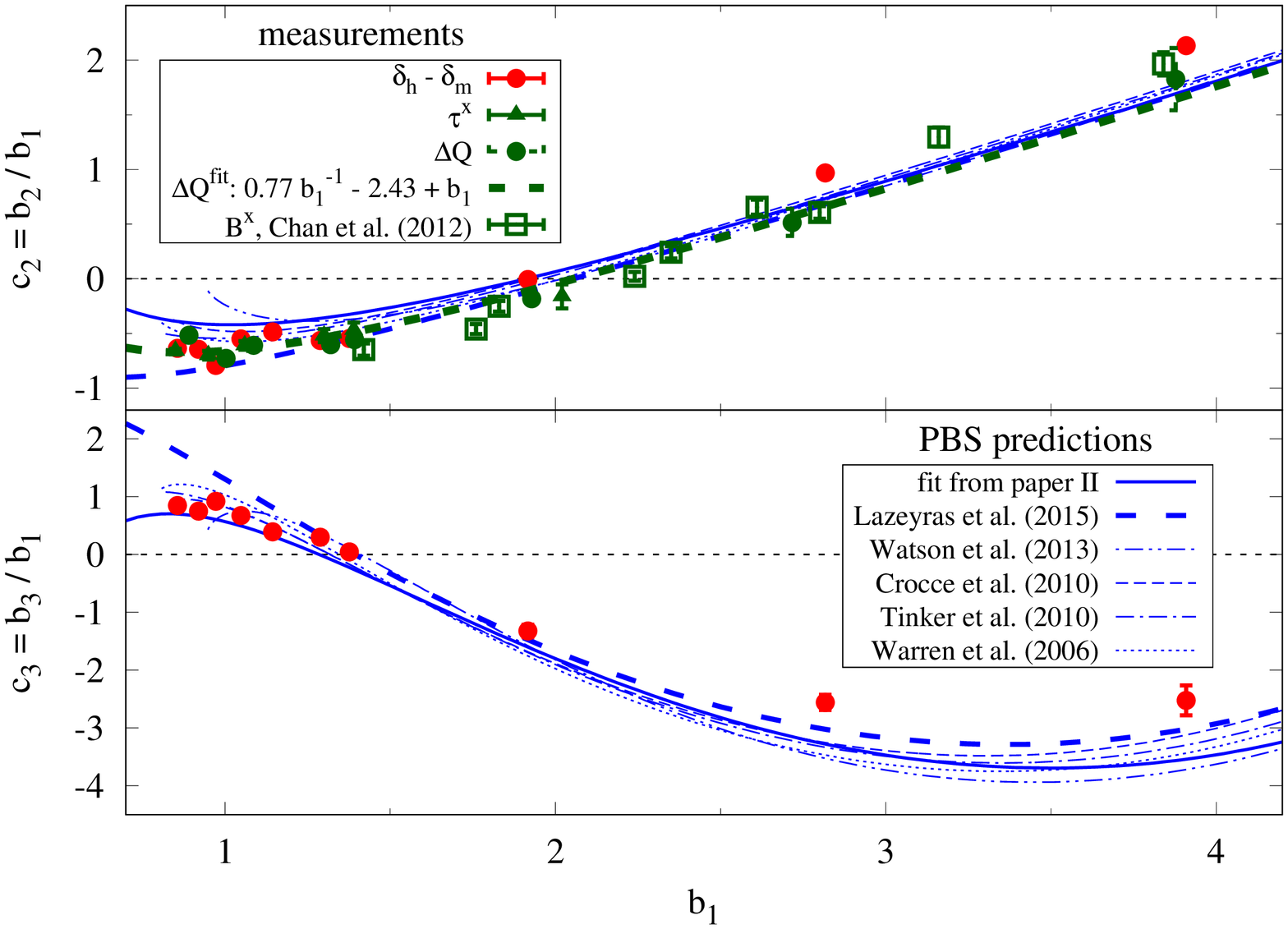}
	\caption{Second- and third-order bias parameters versus the linear bias parameter $b_1$
	(top and bottom panel respectively). Results from $\Delta Q$ and $\tau^{\times}$), $\delta_h-\delta_m$
	as well as PBS predictions from \paperIII\ are the same as those shown in Fig. 
	\ref{fig:bias_comparison}. In addition we show PBS prediction based on mass 
	function fits from the literature. Fits to the $b_1-c_1$ and $b_1-c_3$ relations 
	from separate universe simulations in combination with the PBS model, given by \citet{Lazeyras15}
	are shown as long dashed blue lines.}
	\label{fig:b1-b2-b3}
\end{figure*}

In Fig. \ref{fig:b1-b2-b3} we show the linear bias versus the quadratic and 
third-order bias parameters from the different measurements together with the 
PBS predictions for redshift $z=0.0$ and $0.5$.

\rev{
The PBS predictions labeled as \citet{Crocce10} are derived from FoF mass function fits of these authors to nested boxes
runs from the MICE simulation suite which provides a  higher mass resolution in the low mass range. The PBS
predictions from the mass functions from \citet{warren06}, \citet{Tinker10} and \citet{Watson13} are based on
simulation with cosmologies different from the one of MICE. In addition, \citet{Tinker10} defined haloes as spherical over-densities instead of FoF groups. These difference can lead to deviations in the mass function (see \paperIII) and therefore
contribute do the differences which we see among the various PBS bias
predictions for the $b_N(b_1)$ relation in Fig. \ref{fig:b1-b2-b3}.
Note that the BPS results from \citet{Lazeyras15}, which also use spherical over-densities as haloes, have not been predicted
from the mass function but were directly measured using separate universe simulations.}
It is interesting to note that the $b_2(b_1)$ relation from \citet{Lazeyras15} agrees best with our measurements derived from the $\Delta Q$
method, since the latter also delivers very accurate measurements of  the linear bias. This suggests that also the $c_2$ measurements and
hence the $b_2-c_2$ relations from these two approaches are reliable. Furthermore we include bias measurements from \citet{chan12},
which are based on the Bispectrum and find a good agreement with our results.

Overall we find consistent results from the different measurements and predictions from simulations with different cosmologies
analysed at different redshifts. This indicates a roughly universal behaviour between the linear and higher-order bias 
parameters.
\rev{In \paperIII\ we derived an analytic expression for $b_N(b_1)$ from the PBS bias parameters based on the
\citet{PS74} mass function, which has the form
\begin{equation}
  b_N = \sum^{N=2}_{n=0} \alpha_n b^n.
   \label{eq:b1-bn}
\end{equation}
Since this expression is independent from the peak-height, we also do not expect it to depend on how halo masses are defined.
The $b_N(b_1)$ predictions from \paperIII\, shown in Fig. \ref{fig:b1-b2-b3}, have been fitted to PBS predictions for $b_2(b_1)$ and $b_3(b_1)$
which are based on the the Tinker mass function model, fitted to the MICE-GC mass function at z = 0.0 in the same article.}
Fitting the polynomial from equation (\ref{eq:b1-bn}) to measurements from $\Delta Q$ we find $(\alpha_0,\alpha_1,\alpha_2) = (0.77,-2.43,1)$.

\section{Summary and Conclusion}\label{sec:conclusion}


This analysis is the last part of a series of articles on the accuracy of bias parameters derived from
the MICE Grand Challange (MICE-GC) simulation using clustering statistics and peak-background
split (referred to as PBS) predictions \citep[][\paperII, \paperIII]{HBG15-1}.
In the present analysis we  studied bias  parameters derived from the relation between matter
and halo density contrasts ($\delta_h,\delta_m$ respectively) as an additional method for measuring bias.
These measurements are compared to a selection of our most robust previous results \rev{using the same
four mass samples M0-M3 at redshift $0.0$ and $0.5$. Thanks to the large volume and resolution of the
MICE-GC simulation these samples span a large mass range from between roughly $10^{12}$ 
and $10^{15}$ \Msun, which corresponds to Milky Way like haloes and massive galaxy clusters respectively.}

Our previous have been derived from
two-point halo-matter cross-correlations ($\xi^{\times}$),
a combination of three-point halo-matter auto- and cross-correlations ($\Delta Q$) 
as well as a combination of the halo-matter cross-skewness and cross-correlators ($\tau^\times$).
We thereby employ leading-order modelling of clustering statistics, at which the 
linear bias parameters from these estimators are not affected by non-local contributions to the bias model.
We therefore obtain the linear bias from $\xi^{\times}$ and the linear and 
quadratic bias from $\Delta Q$ and $\tau^\times$.
The PBS predictions are based on MICE-GC mass function fits from \paperIII, using different 
mass function models, while we study bias parameters up to 
order three, for which we have corresponding measurements from the $\delta_h-\delta_m$ relation for validation.

We studied in this work bias measurements from the $\delta_h-\delta_m$ relation in two ways.
The more common method is to fit a polynomial to the $\delta_h-\delta_m$ relation, measured 
in the simulation. Alternatively we explore bias measurements obtained from 
the probability distribution function (referred to as PDF) of $\delta_h$ and $\delta_m$. The latter 
method has the advantage that it can be directly applied to observations, since the PDF 
of matter density contrasts can be modelled with theory or simulations
\citep[][]{bel2016,DiPorto2015, mar05, bernardeau02}. However, it has not been tested so far 
how accurately the bias parameters can be determined with this method.

The results of our bias comparison are summarized in Fig. \ref{fig:bias_comparison}.
In the case of the linear bias $b_1$ we consider results from $\xi^\times$ ($b_\xi^\times$) as the most reliable
since non-linear local  and non-local term can safely be neglected. Furthermore, $b_{\xi}$ is highly
relevant for cosmology since it is weakening constraints on cosmological parameters inferred from $\xi$ due
to its degeneracy with the growth of matter fluctuations.

Linear bias measurements and predictions
from all other methods are in a $\sim 5$ percent agreement with $b_\xi^\times$, while $\Delta Q$ delivers the most accurate
measurements with an overall percent level accuracy. Standard measurements from the $\delta_h-\delta_m$ 
relation are in a slightly better overall agreement with $b_\xi^\times$ than those from  the $\delta$PDF-method.
The strongest deviations from $b_\xi^\times$ shows the linear bias predictions from the different PBS  model.
These predictions are consistently up to $7$ percent below $b_\xi^\times$ at high masses as reported in the
literature \citep[e.g.][]{M&G11, Pollack12}. \rev{The fact that we find similar results for different mass function models
indicates shortcomings in the standard PBS modelling, i.e. the assumption of a constant matter density threshold for gravitational collapse
\citep{Paranjape13b, Paranjape13a, Lazeyras15}.} We do not find a clear change of the variation between the different
measurements and predictions with mass or redshift.

In the case of the quadratic bias $c_2 \equiv b_2/b_1$ we find 
consistent results from the different measurements and predictions as $c_2$ increases from 
negative values of $\gtrsim -0.5$ at low halo masses ($\sim 10^{12} M_{\odot}/h$) to positive values
of up to $\sim 2$  at high masses ($\gtrsim 10^{14} M_{\odot}/h$).
However, the variation between the different results tends to be larger than in the case of 
$b_1$. In the case of the measurements this effect is presumably caused by the strong assumptions such
as the validity of tree-level perturbation theory ($\Delta Q$ and $\tau^\times$), Poissonian shot-noise
($\tau^\times$), or a local deterministic bias model ($\delta_h - \delta_m$). We therefore do not consider any
of these measurements to be sufficiently reliable for being a reference, such as the results from $\xi^\times$ 
in the case of the linear bias. However, the fact that the $\Delta Q$ method delivers highly accurate measurements of
the linear bias suggests that also the quadratic bias is measured reliably with this approach.
We find the PBS predictions for $c_2$ to be consistently above (below) all measurements in the low (high)
mass range as they show overall weaker mass dependence. The latter finding lines up with reports from
\citet{M&G11} and \citet{Pollack12}.

Both measurements of the third-order bias $c_3\equiv b_3/b_1$ from the $\delta_h-\delta_m$ relation 
agree overall mutually at the $1\sigma$ level. These measurements allow for 
a validation of the corresponding PBS predictions. Results from both methods are similar as $c_3$ 
is positive below unity  in the low mass range  and decreases to negative values of down to
$\sim -3$ in the high masses range. However, deviations are significant, as the PBS predictions tend to be
below the measurements at low halo masses while results based on different mass function fits are consistent
with each other.

We use our various linear and non-linear bias measurements for validating the universal polynomial
relation between linear and non-linear bias ($b_2(b_1)$ and $b_3(b_1)$), which we deduced in
\paperIII\ from PBS predictions
\rev{based on the \citet{PS74} mass function. Since this expression is independent of the peak-height
we do not expect a strong dependence on halo mass definition.}
Our measurements show an overall agreement with the universal behaviour predicted by the PBS model.
Furthermore they agree with results from the literature derived via the Bispectrum in Fourier space or the
separate universe approach from simulations with cosmologies different to the one of MICE
\citep[i.e.][]{chan12,Lazeyras15}.
We fit a second-order polynomial to $b_2(b_1)$ measurements from
$\Delta Q$, which we consider as the most reliable $c_2$ estimator as mentioned above.

Such a universal relation between  linear and non-linear bias can be useful for reducing errors on the
linear bias and the growth from clustering analysis when the latter is affected by $c_2$, for instance in 
the case of three-point correlations or two-point correlations at small scales. 

For applying universal polynomial relations between bias parameters in the analysis of galaxy 
surveys it would be interesting to show that their universality also holds for halo samples,
which are selected by galaxy properties, such as luminosity and colour instead of halo 
mass. A correlation between the linear and quadratic bias from the 3pc and Bispectra has been reported
by for SDSS galaxy samples and mock HOD catalogues by \citet{Kayo04} and 
\citet{Nishimichi07} and compared with PBS predictions. The limited accuracy 
of their measured $b_1-c_2$ relations could be strongly improved using the 
methods studied in the present analysis in combination with data from upcoming 
galaxy surveys.


\FloatBarrier
\section*{Acknowledgements}

Funding for this project was partially provided by the Spanish Ministerio de Ciencia e Innovacion (MICINN), project AYA2009-13936,
AYA2012-39559 and AYA2015-71825,
Consolider-Ingenio CSD2007- 00060, European Commission Marie Curie Initial Training Network CosmoComp (PITN-GA-2009-238356)
and research project 2014 SGR 1378 from Generalitat de Catalunya.
KH is supported by beca FI from Generalitat de Catalunya and MINECO, project ESP2013-48274-C3-1-P.
He also acknowledges the Centro de Ciencias de Benasque Pedro Pascual where parts of the
analysis were done.
The MICE simulations have been developed by the MICE collaboration at the MareNostrum
supercomputer (BSC-CNS) thanks to grants AECT-2006-2-0011 through AECT-2010-1-0007.
Data products have been stored at the Port d'Informació Científica (PIC).

We thank Martin Crocce, Pablo Fosalba, Francisco Castander and Roman
Scoccimarro for interesting and useful comments.

\bibliographystyle{mnbst}
\bibliography{hbg.bib}


\end{document}